\documentclass[aps,prd,notitlepage,nofootinbib,superscriptaddress,twocolumn]{revtex4-1}

\usepackage[utf8]{inputenc}
\usepackage{amsmath}
\usepackage{amssymb}
\usepackage{amsfonts}
\usepackage{newtxtext,newtxmath}
\usepackage{bm}
\usepackage{graphicx}
\usepackage[usenames,dvipsnames]{xcolor}
\usepackage{color}
\usepackage[colorlinks=true,linkcolor=blue,urlcolor=blue,citecolor=blue]{hyperref}

\usepackage{orcidlink}

\begin{document}

\title{Can average speed of sound and thermodynamic response functions signal the  exotic phases in neutron star cores?}

\author{Suman Pal,\orcidlink{0009-0000-5944-4261}}
\email{sumanvecc@gmail.com}
\affiliation{Physics Group, Variable Energy Cyclotron Centre, 1/AF Bidhan Nagar, Kolkata 700064, India}
\affiliation{Homi Bhabha National Institute, Training School Complex, Anushakti Nagar, Mumbai 400085, India} 
\author{Gargi Chaudhuri,\orcidlink{0000-0002-8913-0658}}
\email{gargi@vecc.gov.in}
\affiliation{Physics Group, Variable Energy Cyclotron Centre, 1/AF Bidhan Nagar, Kolkata 700064, India}
\affiliation{Homi Bhabha National Institute, Training School Complex, Anushakti Nagar, Mumbai 400085, India}

\begin{abstract}
The speed of sound in dense nuclear matter is crucial for understanding neutron star structure and constraining the EOS. We have discussed in details the decomposition of speed of sound  via  the average speed of sound and its logarithmic derivative and have connected it to the other two decomposition schemes via slope and curvature of the energy per particle or through the normalized trace anomaly and its derivative. These thermodynamic variables provide important diagnostic tools for the composition of the inner core of the compact stars. We discuss a new method of understanding  phase transition and the microphysics of dense matter through the thermodynamic response functions like isothermal compressibility, baryon number susceptibility and bulk modulus  in order to distinguish  
between local (sharp interface) and global charge (mixed phase)
neutrality conditions, thereby revealing  the signatures of the phase transition. The corresponding neutron-star mass--radius relations demonstrate that all considered equations of state satisfy current astrophysical constraints, while the most massive stable configurations contain either an extended mixed phase or a quark core depending on the phase-transition construction.
\end{abstract}

\maketitle


\section{Introduction}
Compact stars serve as exceptional astrophysical laboratories for exploring the properties of dense nuclear matter. Detailed pulsar observations~\cite{Fonseca:2021wxt,Riley:2019yda,Miller:2019cac,Miller:2021qha}, together with gravitational-wave detections~\cite{LIGOScientific:2018cki}, have yielded significant insight and placed important constraints on the nuclear equation of state (EOS). In particular, understanding the detailed structure of the dense matter EOS is crucial for accurately describing the phenomenology of neutron stars.

The speed of sound ($C_s^2$)~\cite{Tan:2021nat,Greif:2018njt,Tews:2018kmu,Fujimoto:2022ohj,Annala:2023cwx,Marczenko:2022jhl,Brandes:2022nxa,PhysRevC.99.035803,Dutra:2015hxa,Zacchi:2015oma,Motornenko:2019arp,Ecker:2022xxj,Sorensen:2021zme,Kojo:2021hqh,Huang:2022mqp,Chiba:2023ftg,Chiba:2024cny,Marczenko:2023txe} is a fundamental quantity inherent to all thermodynamic systems, encapsulating how pressure responds to changes in energy density. In the regime of dense nuclear matter, it plays a pivotal role in determining the internal structure, stability, and maximum mass of neutron stars, as well as in constraining the underlying EOS. 
There is growing interest in the topic of  decomposition of  
the speed of sound. 
One approach \cite{Marczenko:2023txe} characterizes it in terms of the slope ($\alpha$) and curvature ($\beta$) of the energy per particle. 
Another approach \cite{Fujimoto:2022ohj} establishes a direct relationship between the speed of sound and the normalized trace anomaly ($\Delta$), as well as its derivative ($\Delta^{\prime}$). In this framework, $C_s^2$ can be expressed in terms of $\Delta$ and $\Delta^{\prime}$, providing an alternative thermodynamic perspective that links the stiffness of the EOS to deviations from conformal behavior. 

In our previous work\cite{Pal:2025jln}, we had 
 conducted a detailed analysis of the decomposition variables ($\alpha$, $\beta$, $\Delta$ and $\Delta^\prime$) associated with hadronic matter, thoroughly examining their behavior. We also provided a comprehensive explanation of the observed sign change in the parameter $\beta$, highlighting its physical significance and its implications for the underlying microphysics of the system. Building upon this foundation, in the present study, we  have extended our analysis to the  quark star as well as the hybrid star EOS.
Building upon this foundation, in the present work we extend the analysis to pure quark matter and hybrid-star equations of state, allowing us to investigate the behavior of these thermodynamic quantities across the hadron--quark phase transition.

The phenomenon of hadron-quark phase transition is one of the most interesting current topics of research in compact star physics \cite{Most:2018eaw,Alford:2017qgh,Pal:2025chs,Laskos-Patkos:2024otk,Pal:2025skz,Pal:2024afl}. 
A wide range of recent studies has significantly advanced our understanding of phase transitions in dense neutron-star matter. Strong first-order phase transitions and their impact on compact-star structure have been extensively investigated, leading to the possibility of disconnected hybrid-star branches and twin-star configurations~\cite{Alford:2013aca,Alford:2017qgh,Gorda:2022lsk}. 
On the other hand, smooth hadron--quark crossover scenarios have been proposed as an alternative realization of deconfinement at high density, with important consequences for the thermodynamic and transport properties of dense matter~\cite{Huang:2022mqp,Fujimoto:2024ymt}. 
In parallel, Bayesian inference and model-agnostic reconstructions of the dense-matter equation of state have emerged as powerful tools to constrain neutron-star interiors using multimessenger observations~\cite{Xie:2020rwg,Pfaff:2021kse,Albino:2025puc,Chatziioannou:2024jsr}. These approaches allow systematic incorporation of observational data while minimizing model bias in the EOS parametrization.
Thermodynamic quantities such as the speed of sound and the trace anomaly have been widely used as diagnostic tools for identifying changes in the microscopic composition of dense matter~\cite{Cai:2023pkt,Fukushima:2024gmp}. In particular, these quantities provide insights into the degree of conformality and stiffness of the equation of state across different density regimes.
Furthermore, recent works have demonstrated that signatures of phase transitions may also manifest in gravitational-wave signals from binary neutron-star mergers and their post-merger evolution~\cite{Most:2018eaw,Fujimoto:2022xhv,Fujimoto:2024ymt,Mondal:2023gbf}. These developments establish a direct connection between microphysical phase structure and observable astrophysical signals.

In order to model the transition between hadronic and quark phases in neutron star interiors, researchers commonly employ either the Gibbs or the Maxwell approach, depending on the value of the surface tension at the hadron-quark interface \cite{Endo:2005zt}.
For this purpose, we employ both Maxwell and Gibbs constructions—capturing sharp first-order phase transitions characterized by local charge neutrality and constant pressure, as well as mixed-phase coexistence under global charge neutrality. This combined approach offers a framework to investigate the intricate thermodynamics and phase structure inherent in hybrid star equations of state.

Although the quantitative results depend on the chosen EOS, bag constant, and vector coupling, the main qualitative features of the speed-of-sound decomposition remain robust across the variations considered in this work.

In this work, we have proposed another alternative  method of decomposing the speed of sound through the average speed of sound and its logarithmic derivative.
In particular, we focus on the detailed behavior of the different decomposition schemes of speed of sound, $C_s^2$ and also  have established the connection between these different schemes.
 This framework further allows us to investigate the physical origin 
of the approximate universal relation \cite{Saes:2024xmv} between the averaged speed of sound 
and the compactness of compact stars.

We stress that the decomposition framework introduced in this work does not represent a new thermodynamic identity, but rather provides an alternative and physically transparent representation of existing relations. Its utility lies in connecting different formulations of the speed of sound decomposition within a unified framework and in facilitating comparisons across different phases of dense matter.

We also propose  new signatures of the hadron--quark phase transition by analyzing the differences between the Maxwell and Gibbs constructions through thermodynamic response functions. In this work we have analyzed the behavior of baryon number susceptibility, isothermal compressibility and bulk modulus. These response functions can easily distinguish between local and global charge neutrality conditions and thereby act as a signature of the phase transition.
 We have linked the response functions to the speed of sound decomposition variables.

This analysis highlights clear differences that arise depending on the construction scheme used. Our study demonstrates that the contrasting behavior of decomposition variables and response functions between the MC and GC scenarios may serve as a sensitive probe of the phase transition. 
All thermodynamic quantities are interconnected through the speed of sound. Therefore, the behavior of $C_s^2$ provides a valuable diagnostic tool to distinguish between competing descriptions of the phase transition and to shed light on the micro-physical properties governing hybrid star matter.

While previous studies have explored the decomposition of the speed of sound and related thermodynamic quantities in hadronic matter, much less is known about how these different representations behave across the hadron--quark transition and within hybrid-star equations of state. In particular, it remains important to understand whether different decomposition schemes provide complementary insight into the phase structure of dense matter and how they are connected to thermodynamic response functions.

Motivated by these questions, we investigate pure hadronic, pure quark, and hybrid equations of state constructed using both Maxwell and Gibbs approaches. We analyze the behavior of the speed of sound through the slope--curvature decomposition, the trace-anomaly representation, and a decomposition based on the pressure-to-energy-density ratio and its logarithmic derivative. We further examine the baryon number susceptibility, isothermal compressibility, and bulk modulus, and explore their connection to the underlying sound-speed behavior. This unified analysis allows us to compare different thermodynamic descriptions of dense matter and to identify characteristic differences between Maxwell and Gibbs constructions of the hadron--quark phase transition.

\section{Thermodynamic variables.}
\label{sec:forma_thermo_vari}
In this section, we review the analysis of the speed of sound and other key physical quantities. The EOS provides a fundamental link between pressure ($P$) and energy density ($\varepsilon$). 

\subsection{Speed of sound }
As introduced earlier in previous studies\cite{Pal:2025jln}, the decomposition of $C_s^2$ can be executed using two complementary methods as described below:
By applying thermodynamic identities--namely, $d\varepsilon=\mu d\rho$ and $P=\rho^2\frac{d}{d\rho}\left(\frac{\varepsilon}{\rho}\right)$
—the speed of sound can be expressed as the sum of two components \cite{Marczenko:2023txe,Pal:2025jln}:
\begin{equation} \label{eq:cs2_decompose}
    C_s^2=\frac{1}{\mu}\left(\frac{dP}{d\rho}\right)=2 \frac{\rho}{\mu}\frac{d}{d \rho}\left(\frac{\varepsilon}{\rho}\right)+\frac{\rho^2}{\mu}\frac{d^2}{d \rho^2}\left(\frac{\varepsilon}{\rho}\right)=\alpha+\beta
\end{equation}
the 1st term  on the right hand side being proportional to the first derivative (slope) of energy per baryon and the 2nd one being proportional to the 2nd derivative (curvature) of energy per baryon . 

Another important quantity that relates the energy density and pressure in a star is the polytropic index ($\gamma$) \cite{glendenning2012compact,Marczenko:2023txe}. It can be expressed as the logarithmic derivative of pressure and energy density as follows:
\begin{equation} \label{eq:poly_index}
    \gamma=\frac{d~log~P}{d~log~\varepsilon}=\frac{\varepsilon}{P}C_s^2 
\end{equation} 
We can relate the parameters $\alpha$ and $\beta$ to the speed of sound and polytropic index as indicated below. 
\begin{equation}
    \alpha=\frac{2 P}{P+\varepsilon}=\frac{2 C_s^2}{C_s^2+\gamma},~~\beta=C_s^2-\alpha
\end{equation} 

An alternative approach\cite{Fujimoto:2022ohj} to the decomposition involves the normalized trace anomaly ($\Delta$), which has recently attracted considerable interest in neutron star research. The trace anomaly, normalized by the energy density, is defined as:  
\begin{equation}
    \Delta = \frac{1}{3} - \frac{P}{\varepsilon}.
\end{equation}  
A particularly insightful quantity related to $\Delta$ is its derivative with respect to energy density, denoted as $\Delta'$. Using these, the speed of sound can be expressed as~:  
\begin{equation}
    C_s^2 = \frac{1}{3} - \Delta - \Delta',
\end{equation}  
where $\Delta' = \varepsilon \frac{d\Delta}{d\varepsilon}$.
\begin{figure}[htbp]
\centering
\includegraphics[width=0.22\textwidth]{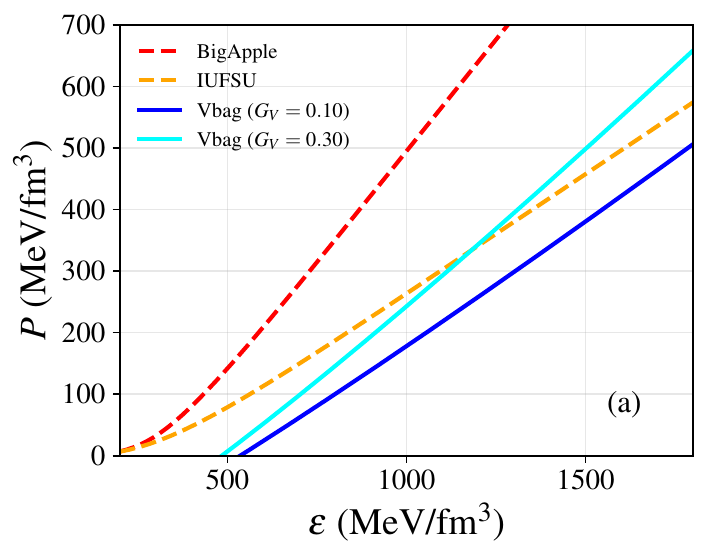}
\includegraphics[width=0.22\textwidth]{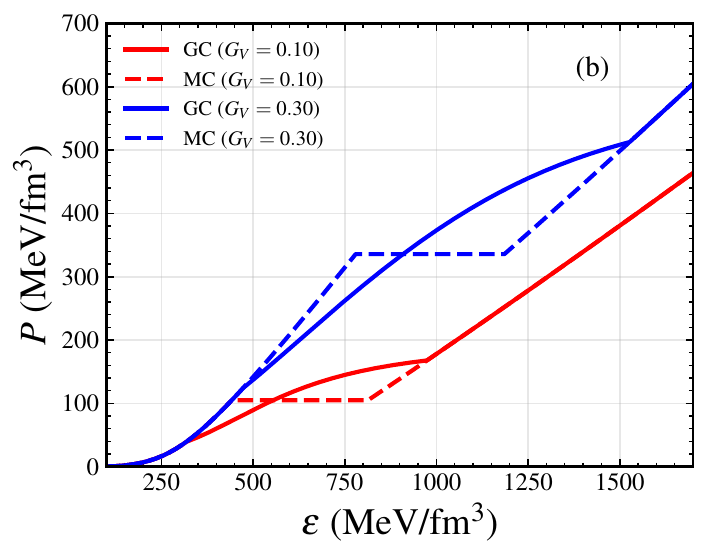}
\caption{Description of EOS (a) for hadron and quark matter (b) and for hybrid case showing the transition from hadronic to quark matter.}
\label{fig:1}
\end{figure}

This formulation provides a valuable thermodynamic perspective that links deviations from conformality to the stiffness of dense matter. 
Qualitative insights into the EOS can be gained by examining the ratio $W = \frac{P}{\varepsilon}$ and its derivative with respect to $\varepsilon$, we can also call $W $ as a average speed of sound \cite{Marczenko:2024uit} as:
$ <C_s^2>=\frac{\int_0^\varepsilon C_s^2 d\bar{\varepsilon}}{\int_0^{\varepsilon} d\bar{\varepsilon}}=\frac{P}{\varepsilon}=W
$. We propose the decomposition of the speed of sound as sum of average speed of sound W ($<C_s^2>$) and its logarithmic derivative with respect to energy density $W^{\prime}$ ($\varepsilon \frac{dW}{d\varepsilon}$)
\begin{equation}
    C_s^2 = W + W^{\prime},
\end{equation} 
 The different decomposition parameters are related to each other by the following equations:
 $  \alpha=\frac{2W}{1+W},~~\beta = W^{\prime}+\frac{W(1-W)}{1+W}$ and $  \Delta =\frac{1}{3}-W,~~\Delta^{\prime}=-W^{\prime}.$

We note that the identification of $W = P/\varepsilon$ with an average speed of sound implicitly assumes that the zero-pressure point corresponds to $\varepsilon=0$. While this is appropriate for hadronic equations of state, it is not valid for self-bound systems such as quark matter, where $\varepsilon(P=0)=\varepsilon_0 \neq 0$. In such cases, a generalized definition of the averaged speed of sound must be employed. For completeness, we provide this generalized formulation in Appendix~\ref{app:selfbound}.

The usefulness of this representation lies in its interpretational transparency. In particular, expressing the speed of sound in terms of $W = P/\varepsilon$ and its logarithmic derivative $W'$ provides a direct way to disentangle the contributions from the magnitude of the pressure-to-energy-density ratio and its variation with energy density. This form enables a more transparent comparison between different equations of state and helps to identify qualitative changes in the thermodynamic response, especially in the presence of phase transitions.

In this sense, the proposed decomposition should be regarded as a complementary reorganization of existing thermodynamic relations, offering additional physical insight rather than a fundamentally new formulation.

The range of these decomposition parameters $\alpha$, $W$ and  the normalized trace anomaly  $\Delta$ can be obtained from the  conditions of thermodynamic stability and  the causality limits which impose the restrictions $P>0$ and $P \leq \varepsilon$, respectively.  Consequently, the variables $\alpha$ and $W$  is  constrained to the range $0 \leq \alpha,W < 1$ and $\Delta$ is constrained to the range $-2/3 \leq \Delta < 1/3$.

\begin{figure*}[htbp]
\centering
\includegraphics[width=0.95\textwidth]{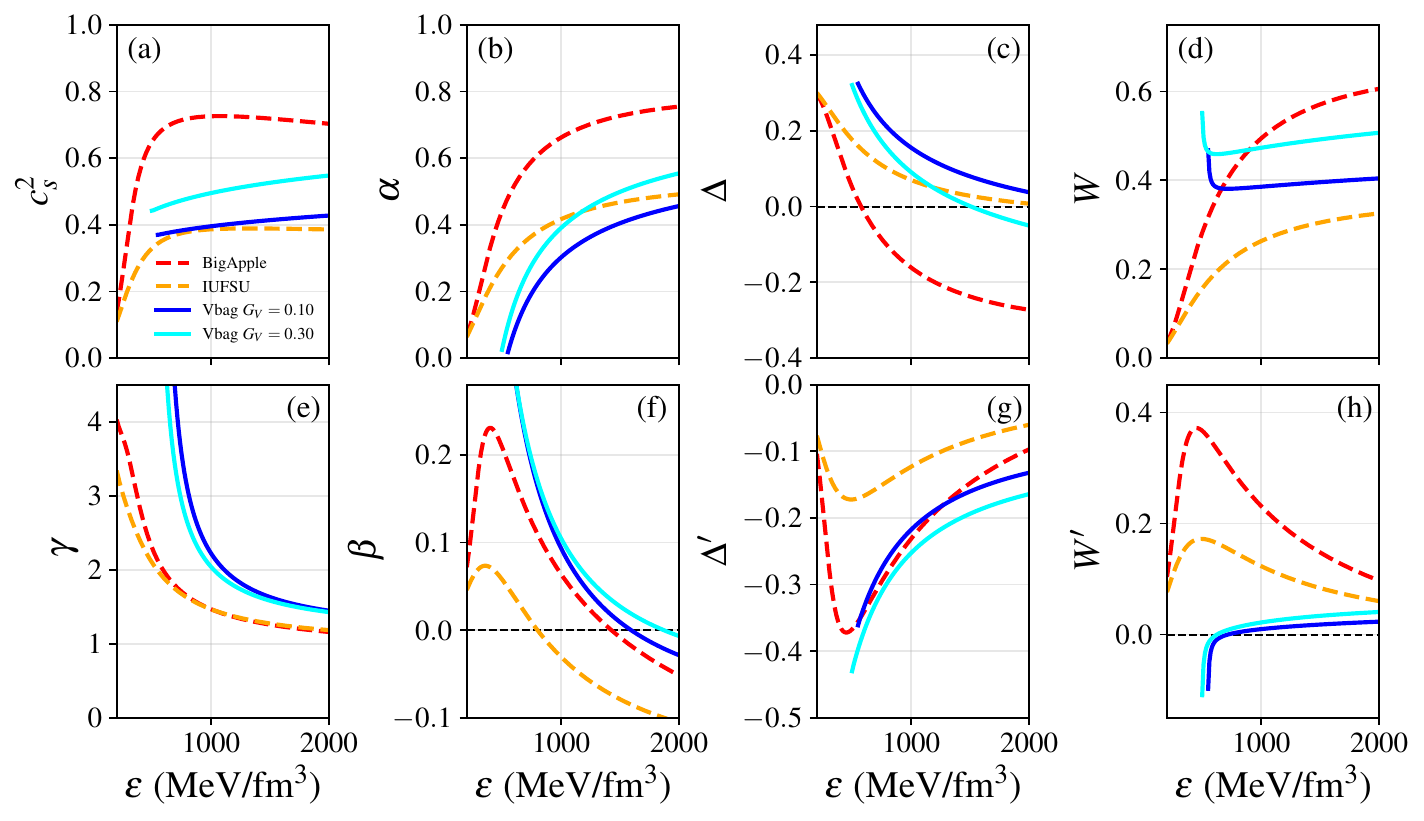}
\caption{Description of decomposition of speed of sound for pure hadron matter and pure quark matter.}
\label{fig:2}
\end{figure*}

\subsection{Response functions}\label{subsec:forma_response_func}

 The macroscopic properties of a system are governed by thermodynamic potentials, 
such as the grand potential per unit volume ($\Omega$). The first derivatives of these potentials with respect to their 
natural variables (e.g., baryon chemical potential $\mu_B$) yield the corresponding conjugate variables 
(e.g., particle baryon density $\rho_B = -\partial \Omega / \partial \mu_B$). 
The second derivatives of the potentials, which quantify how these conjugate variables respond to 
changes in the control parameters, are the thermodynamic response functions. 
In our cold neutron star system ($T=0$), the quantities of interest are the isothermal compressibility, bulk modulus and baryon number susceptibility. Their behavior can throw light on the properties  of quark-hadron phase transition in the neutron star core.
An important thermodynamic quantity is the baryon number susceptibility, denoted as 
\begin{equation} \label{eq:chi_n}
    \chi_N = \left( \frac{\partial \rho_B}{\partial \mu_B} \right) = -\frac{\partial^2 \Omega}{\partial \mu_B^2}=\frac{\rho_B}{\mu_B C_s^2}.
\end{equation}
The isothermal compressibility $K_T$ is a key thermodynamic parameter that quantifies the stiffness of the EOS and plays a critical role in analyzing the phase structure of the system. The speed of sound, baryon number susceptibility and enthalpy density ($h=P+\varepsilon$) can be related to $K_T$ through the following expression:
\begin{equation}\label{eq:K_T}
    K_T = \frac{1}{\rho_B} \left( \frac{\partial \rho_B}{\partial P} \right)_T = \frac{1}{(P + \varepsilon) C_s^2}=\frac{1}{h~C_s^2}=\frac{\chi_N}{\rho_B^2}.
\end{equation}

$K_T$ can be expressed as a second derivative of the thermodynamic potential 
\begin{equation}
    K_T=\frac{1}{\rho_B^2} \left( \frac{\partial \rho_B}{\partial \mu_B} \right)_T =-\frac{1}{\rho_B^2}\frac{\partial^2 \Omega}{\partial \mu_B^2}
\end{equation}
We can also define the bulk modulus ($K_B$ ) as inverse of $K_T$ ; can be expressed as second derivative of the energy density and can be related with speed  of sound and baryon number susceptibility :
\begin{equation}
    K_B=\rho_B  \left( \frac{\partial P}{\partial \rho_B} \right)_T =\rho_B^2\frac{\partial^2 \varepsilon}{\partial \rho_B^2} = (P+\varepsilon)C_s^2 =h~C_s^2= \frac{\rho_B^2}{\chi_N}
\end{equation}
We can relate the response functions with the decomposition variables of $C_s^2$
as:
{\small
\begin{eqnarray} \label{eq:decomp_response_func}
   \chi_N &=&\frac{3\rho_B}{\mu_B(1-3\Delta-3\Delta^{\prime})}=\frac{\rho_B}{\mu_B(\alpha+\beta)}= \frac{\rho_B}{\mu_B(W+W^\prime)}~~~~~~~~\\ \nonumber
   K_B&=& h~C_s^2= h~(\alpha+\beta)= h~(\frac{1}{3}-\Delta-\Delta^{\prime})=\frac{1}{K_T}.\\ \nonumber 
\end{eqnarray}
}

The response functions introduced in Eqs.~\eqref{eq:chi_n}--\eqref{eq:decomp_response_func} are related to the speed of sound and other thermodynamic quantities through exact thermodynamic identities. Consequently, they do not constitute independent observables, and their behavior reflects the same underlying equation of state. In particular, the divergent or vanishing behavior observed in the Maxwell construction originates from the constant-pressure region associated with the first-order phase transition and is therefore directly connected to the corresponding behavior of $C_s^2$.
Nevertheless, these quantities provide useful complementary diagnostics of the thermodynamic response of the system. Since they are related to second derivatives of the thermodynamic potential, they quantify the susceptibility of the system to external perturbations and often amplify rapid changes in the equation of state. As a result, features associated with the phase transition, especially the distinction between Gibbs and Maxwell constructions, can become more transparent when viewed through the behavior of the baryon number susceptibility, isothermal compressibility, and bulk modulus. Throughout this work, these response functions are therefore employed as alternative thermodynamic representations of the same underlying physics rather than as independent signatures of the phase transition.

\section{Results}
\label{sec:result}
\subsection{Equation of state}

 We  have considered  EOS of pure hadron star, pure quark star as well as that of a hybrid for calculation of the thermodynamic quantities. The hadronic phase is described using a relativistic mean field (RMF) model \cite{Horowitz:2000xj,Fattoyev:2010mx,Fattoyev:2020cws}, consisting of nucleons ($p$ and $n$) and
leptons ($e$), while the quark phase is modeled with the MIT bag model with repulsive vector interactions. \cite{glendenning2012compact,Pal:2023quk}.
For the mixed phase in the hybrid star, we use both Maxwell (local charge neutrality ) and Gibbs (global charge neutrality) constructions. The details are discussed in Refs.\cite{Glendenning:1992vb, Glendenning:2001pe,Bhattacharyya:2009fg,Podder:2025qaz,Pal:2026cji}.
It is now well established that different phases of dense matter may leave observable imprints on macroscopic neutron-star properties. Therefore, a central challenge in modern compact-star physics is to construct robust and physically transparent diagnostics that can distinguish between different microscopic descriptions of dense matter.
In this context, it is important to distinguish between model-independent thermodynamic relations and model-dependent realizations of the equation of state. While thermodynamic identities, such as those governing phase equilibrium and the relations among thermodynamic variables, are universal, their quantitative implementation and the resulting physical observables depend on the underlying microscopic framework. In the present work, we employ two representative hadronic RMF models together with a MIT bag-model description of quark matter. Although the quantitative results depend on the chosen EOS, the main qualitative features of the speed-of-sound decomposition remain robust across the variations considered in this work. The decomposition itself follows directly from thermodynamic identities and is therefore model independent, whereas the detailed quantitative behavior of the decomposed quantities reflects the specific EOS employed.

In Fig.~\ref{fig:1}, we show the EOS of pure hadron star with RMF based on BigApple \cite{Fattoyev:2020cws} and IUFSU \cite{Fattoyev:2010mx}) and and pure quark star with MIT Bag model ( bag constant $B_0^{1/4}$=180 MeV ) with vector interaction($G_V$ = 0.1 fm$^2$ and 0.3 fm$^2$)  in the left figure and the hybrid EOS with both MC and GC for the hadron to quark phase transition in the right one. 

\subsection{Decomposition of speed of sound}\label{subsec:results_decompose}
In Fig.~\ref{fig:2}, we display the speed of sound and its decomposition for pure hadronic and pure quark equations of state. Figure~\ref{fig:2}(a) shows the speed of sound for the two hadronic models and the two quark models, with the corresponding equations of state presented in Fig.~\ref{fig:1}. 

In Fig.~\ref{fig:2}(b), (c), and (d), we plot the first component of the three decomposition schemes of the speed of sound, namely $\alpha$, $\Delta$, and {\color{black}the averaged speed of sound} $W$. {\color{black}For self-bound quark matter, $W$ is evaluated using the generalized definition discussed in Appendix~\ref{app:selfbound}, which properly accounts for the finite energy density at zero pressure.} The variation of $\alpha$ and $W$ is qualitatively similar for both hadronic and quark EOSs, while $\Delta$ exhibits an opposite trend and may become negative for sufficiently stiff EOSs.

Figure~\ref{fig:2}(e) shows the behavior of the polytropic index $\gamma$ for the hadronic and quark models. One can clearly see that the low-energy variation of this quantity differs significantly between the hadronic and quark phases.

The lower panels display the second component of the three decomposition schemes of the speed of sound. The quantity $\beta$ ($C_s^2-\alpha$) exhibits a behavior similar to that of $W^{\prime}$, whereas $\Delta^{\prime}$ shows a qualitatively opposite trend. In the low-energy region, the behavior of $W^{\prime}$, $\Delta^{\prime}$, and $\beta$ differs substantially between hadronic and quark matter. For the hadronic EOSs, $W^{\prime}$ and $\beta$ develop a pronounced maximum, while $\Delta^{\prime}$ exhibits a corresponding minimum. In contrast, the variation of these quantities is considerably smoother for the quark EOSs. This difference can be traced to the rapid increase of the speed of sound in hadronic matter, as seen in Fig.~\ref{fig:2}(a). A stiffer EOS generally leads to a more pronounced extremum in $W^{\prime}$, $\beta$, and $\Delta^{\prime}$.

Another noteworthy feature is that $W^{\prime}$ remains positive for the hadronic EOSs considered here, but may become negative at low energy densities in self-bound quark matter. This behavior arises from the generalized definition of the averaged speed of sound, $W=P/(\varepsilon-\varepsilon_0)$, which is appropriate for systems possessing a finite energy density at zero pressure. In contrast, $\Delta^{\prime}$ remains negative throughout the density range considered, whereas $\beta$ can assume either positive or negative values depending on the stiffness of the EOS.

\begin{figure}[htbp]
\centering
\includegraphics[width=0.45\textwidth]{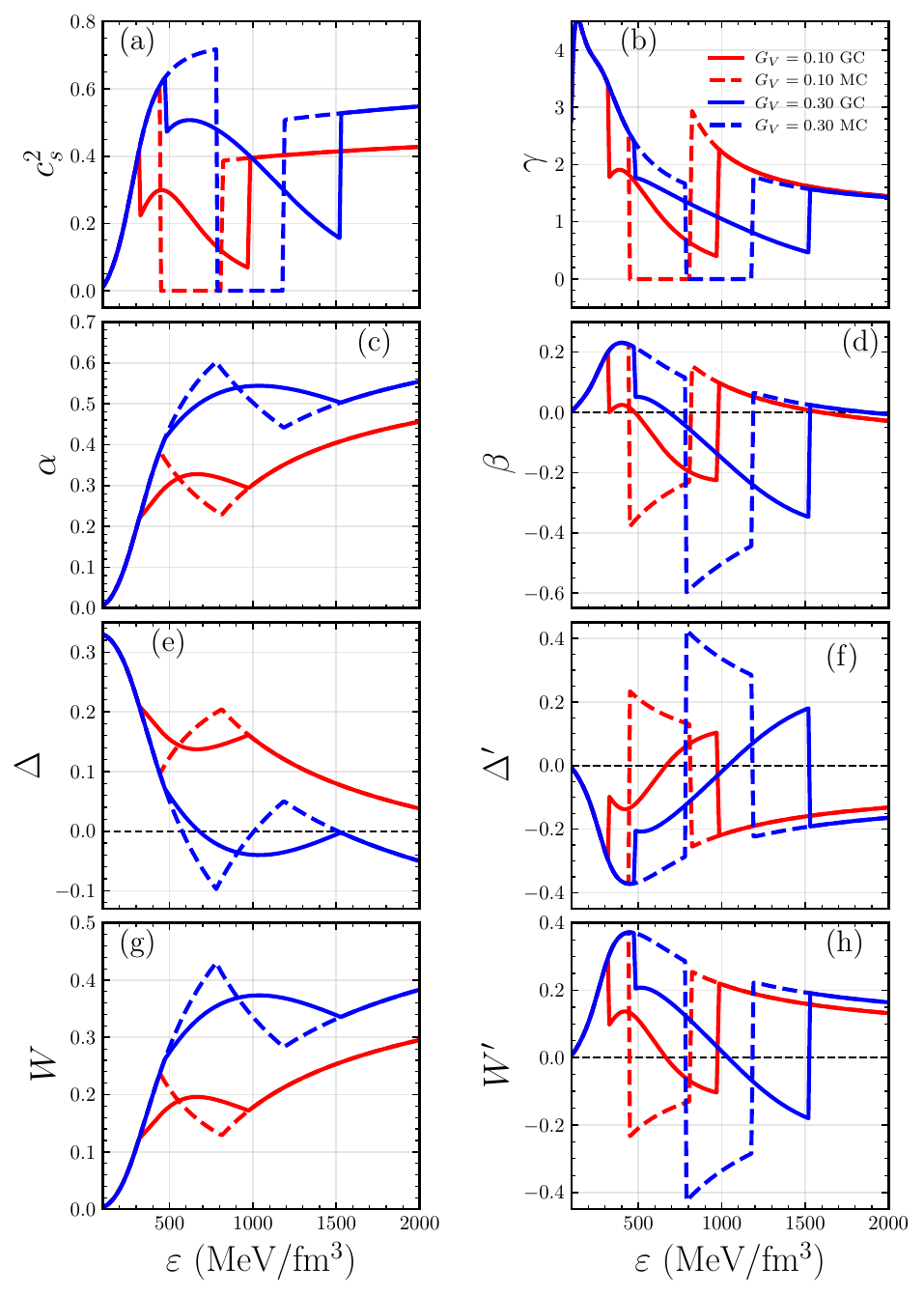}
\caption{Description of decomposition of speed of sound for hybrid star with fixed hadronic EOS (BigApple) with different repulsive interaction both for GC and MC.}
\label{fig:3}
\end{figure}

In Fig.~\ref{fig:3} we repeat all the plots of Fig.~\ref{fig:2} but for hybrid EOS as shown in Fig.~\ref{fig:1}(b). The main motivation of this part is to distinguish between the two mechanisms of phase transition, GC and MC through the different schemes of decomposition of speed of sound. The solid lines correspond to the EOS obtained using the Gibbs construction, while the dashed lines represent the EOS from the Maxwell construction. From Fig.~\ref{fig:3}(a), it is evident that the repulsive interaction expands the range of the mixed phase. 

Moreover, the Gibbs construction gives rise to an extended mixed-phase region, with the transition beginning at lower density and persisting over a broader density interval. In contrast, the Maxwell construction exhibits a first-order transition at constant pressure, characterized by a finite density discontinuity between the hadronic and quark phases.

We find  from Fig.~\ref{fig:3}(a) that in the Maxwell construction (MC) the speed of sound $dP/d\epsilon$ is exactly zero since $P$ remains unchanged. In the Gibbs construction (GC), on the other hand, $C_s^2$ remains finite and even exhibits a peak in the mixed-phase region. At the onset of the mixed phase, the pressure rises faster than the energy density, leading to an increase of $W$ and $\alpha$ and decrease of $\Delta$ with $\epsilon$. Beyond a certain point, however, the growth rate of $P$ becomes slower than that of $\epsilon$, causing $W$ and $\alpha$ to decrease and $\Delta$ to increase as $\epsilon$ continues to increase. This is evident from the Fig.~\ref{fig:3} (c, e and g).

Since the polytropic index $\gamma$ is related to the response of pressure to changes in energy density, the constant-pressure plateau of the Maxwell construction formally implies $dP/d\epsilon=0$, and hence $\gamma=0$ within the coexistence region. This should be interpreted as a consequence of the thermodynamic construction describing phase coexistence between two distinct phases separated by a density discontinuity, rather than as a local property of a homogeneous matter. In contrast, for the Gibbs construction, $C_s^2$ remains finite throughout the mixed-phase region, implying a nonzero value of the polytropic index, as shown in Fig.~\ref{fig:3}(b).

In Fig.~\ref{fig:3}(d, f and h), we present the behavior of $\beta$, $\Delta^{\prime}$ and $W^{\prime}$ respectively  as a function of the energy density.  
In the Maxwell construction, since $C_s^2 = 0$, we obtain the relations 
$\beta = -\alpha$, $\Delta^{\prime} = -\Delta$, and $W^{\prime} = -W$. 
Therefore, the behavior of $\beta$, $\Delta^{\prime}$, and $W^{\prime}$ is entirely determined by the corresponding quantities $\alpha$, $\Delta$, and $W$. 
As $\alpha$ and $W$ are always positive, $\beta$ and $W^{\prime}$ remain strictly negative throughout the mixed phase. In contrast, $\Delta$ can be either positive or negative depending on the stiffness of the EOS (e.g., negative for larger values of the repulsive interaction). Consequently, in the mixed phase, $\Delta^{\prime}$ also take positive values. 
In contrast, under the Gibbs construction $\beta$ and $\Delta^{\prime}$ and $W^{\prime}$ can assume both negative and positive values.

\subsection{Physical interpretation and implications of the decomposition}

We now discuss the physical interpretation of our decomposition. 
The averaged term $W$ encodes the global, integrated stiffness of the equation of state, since it represents the accumulated pressure response up to a given energy density.
For hadronic and hybrid-star EOSs, the averaged quantity is given by $W=P/\varepsilon$. For self-bound quark matter, however, the generalized definition presented in Appendix~\ref{app:selfbound} must be used in order to account for the finite energy density at vanishing pressure.
It determines global stellar properties like mass and radius. In contrast, the 
derivative term reflects how 
rapidly this averaged stiffness changes with increasing energy density, 
thereby capturing the local variation of the equation of state.
Although the speed of sound sensitively measures stiffness, it does not exhibit quasi-universality with compactness -- and this follows directly from its decomposition. It is the derivative term $W'$ that breaks the quasi-universality observed in the averaged speed of sound--compactness relation.
Moreover, a key strength of the present framework lies in its ability to unify previously distinct decomposition schemes within a single, consistent formalism. The parameters $\alpha$, $W$, 
and $\Delta$ are not independent but are related through simple algebraic 
relations. Each parametrization highlights a different aspect: $\alpha$ 
directly probes the density dependence of energy per baryon (through its 
first derivative), $W$ captures the integrated stiffness relevant for 
global stellar properties, and $\Delta$ quantifies the deviation from 
conformal symmetry. Their interrelations ensure consistency while providing complementary insights into the thermodynamic behavior of dense matter.
\color{black}

\begin{figure}[htp] 
	\centering 
    \includegraphics[width=0.45\textwidth]{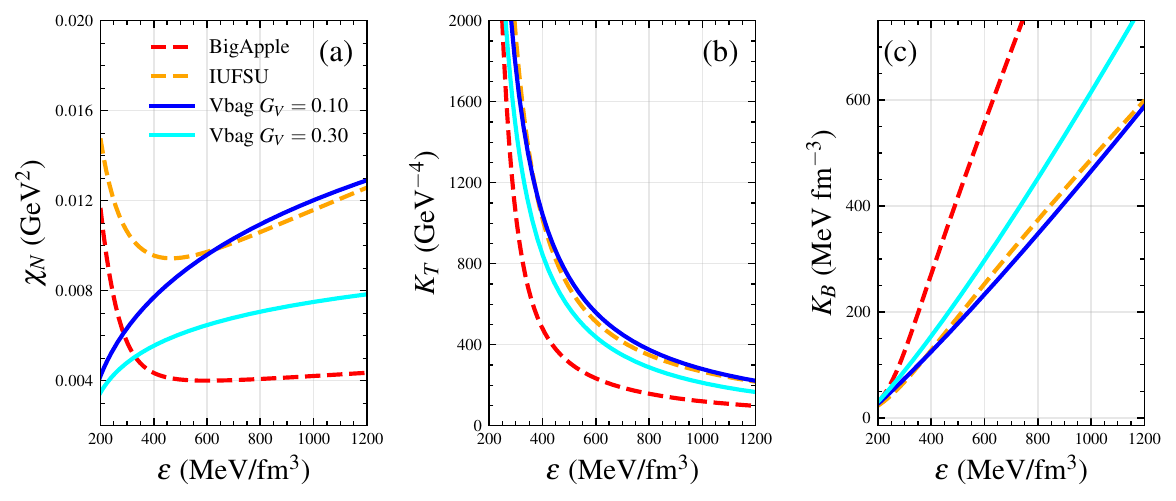}
    \includegraphics[width=0.45\textwidth]{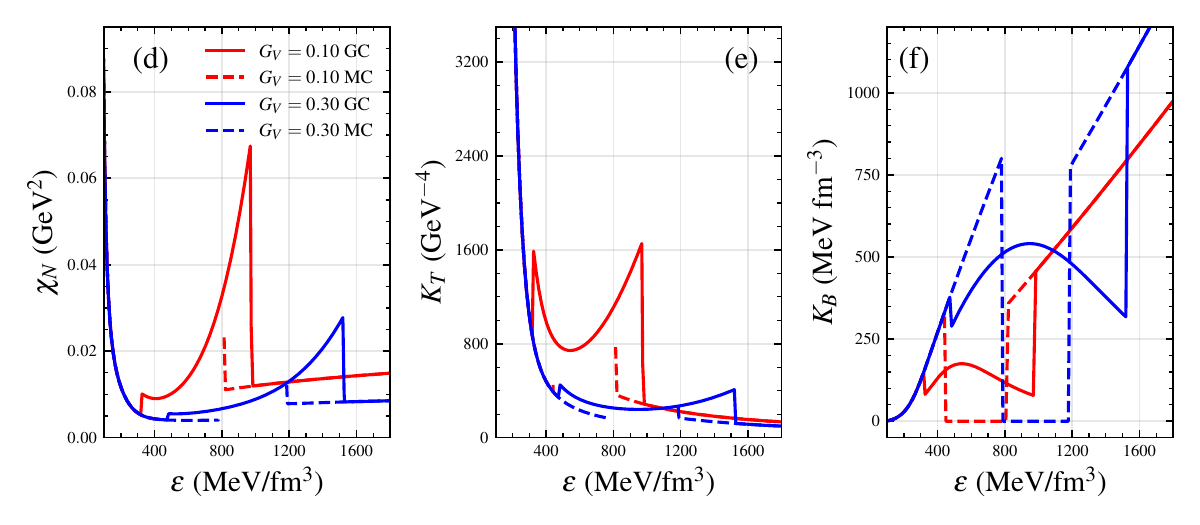}
	\caption{The thermodynamics behavior of the three response functions. Upper panel (a, b and c) represents the variation of the response function for the pure hadronic and quark EOS. Lower panel (d, e and f) represents for the hybrid EOS  with different strength of the repulsive interaction for BigApple EOS both for GC and MC.}   
	\label{fig:4}
\end{figure}


Further insight into the decomposition scheme can be obtained by comparing the pure hadronic, pure quark, and hybrid equations of state.

In the pure hadronic phase, the local speed of sound exceeds its averaged value, leading to positive values of $W^{\prime}$. For self-bound quark matter, however, the generalized definition of the averaged speed of sound can give rise to negative values of $W^{\prime}$ at low energy densities, followed by a sign change at higher densities. In the mixed phase, the averaged speed of sound may become either larger or smaller than the local one, resulting in both positive and negative values of $W^{\prime}$. In particular, the sign change occurring across the mixed phase reflects the competition between the local and averaged stiffness of the equation of state.
This sign reversal signals a structural modification of the stiffness 
evolution and reflects the internal cancellation mechanism responsible 
for the softening of $c_s^2$. Such a feature is not directly visible 
from the sound speed alone, which only encodes the net response.
This behavior of $W'$ naturally connects to other parametrizations of the equation of state. In the slope-curvature description, where $c_s^2 = \alpha + \beta$, the sign change of $W'$ in the mixed phase corresponds to a sign change in $\beta$ when the slope term $\alpha$ crosses $c_s^2$. In both pure and mixed phases, $\alpha$ can exceed $c_s^2$ depending on the softness of the EOS, leading to a corresponding sign change in $\beta$.
In the trace anomaly description, the trace anomaly can take positive or negative values depending on whether the averaged speed of sound is smaller or larger than the conformal speed of sound. 
This provides complementary insight into the approach toward conformality, which is not fully captured by $c_s^2$ alone. Since true conformal behavior requires the simultaneous convergence of both $c_s^2$ and $P/\varepsilon$ to $1/3$, the present decomposition offers a more stringent and direct test of conformal restoration than the local sound speed by itself.

The polytropic Index $\gamma$ is less sensitive to the variation in stiffness both in case of pure hadron star or pure quark star as can be seen from Fig.~\ref{fig:2}(e); on the contrary it is more sensitive to EOS in the mixed phase regime (both GC and MC) as in Fig.~\ref{fig:3}(b). 

\color{black}


\begin{figure}[htp] 
	\centering
         \includegraphics[width=0.45\textwidth]{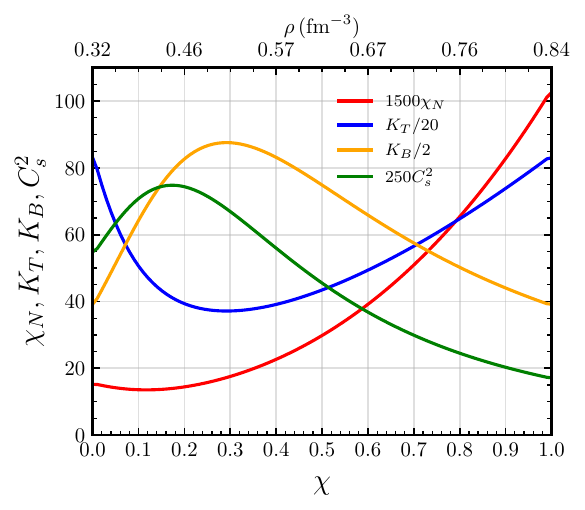}
	\caption{The response functions for the Gibbs construction (BigApple,  $G_V$=0.1 $\text{fm}^2$) with $\chi$(lower x-axis) and with density (upper x-axis).}   
	\label{fig:5}
\end{figure}

\begin{table*}[t]
\caption{Hybrid-star properties and phase structure for different values of $G_V$ with fixed $B^{1/4}=180~\mathrm{MeV}$.}
\setlength{\tabcolsep}{2.0pt}
\begin{tabular}{ccccccccccc}
\hline\hline
$G_v$ & $M_{\rm max}$ ($M_\odot$) & $R_{\rm max}$ (km) & $\rho_c^{\rm max}$ (fm$^{-3}$) & $\epsilon_c^{\rm max}$ (MeV fm$^{-3}$) & $P_c^{\rm max}$ (MeV fm$^{-3}$) & $R_{1.4}$ (km) & $R_{2.0}$ (km) & $\rho_{\rm onset}$ (fm$^{-3}$) & $\rho_{\rm pureQ}$ (fm$^{-3}$) & Phase \\
\hline
\hline
\multicolumn{11}{c}{\textbf{Maxwell Construction}} \\
\hline
0.10 & 2.098 & 13.352 & 0.708 & 805.77 & 104.90 & 13.017 & 13.344 & 0.425 & 0.708 &After density jump \\
0.20 & 2.424 & 13.195 & 0.771 & 941.63 & 190.47 & 13.043 & 13.360 & 0.526 & 0.771 & After density jump \\
0.30 & 2.582 & 12.761 & 0.882 & 1177.29 & 335.73 & 13.037 & 13.356 & 0.655 & 0.882 &  After density jump  \\
\hline
\multicolumn{11}{c}{\textbf{Gibbs Construction}} \\
\hline
0.10 & 1.979 & 12.704 & 0.716 & 812.79 & 153.21 & 13.032 & ---    & 0.328 & 0.835 & Mixed  \\
0.20 & 2.317 & 12.712 & 0.757 & 912.32 & 253.33 & 13.017 & 13.329 & 0.378 & 0.925  & Mixed  \\
0.30 & 2.479 & 12.341 & 0.912 & 1230.82 & 441.47 & 13.032 & 13.354 & 0.448 & 1.075 & Mixed \\
\hline
\end{tabular}
\label{tab:1}
\end{table*}

\subsection{Thermodynamic Response functions for the Hybrid star }\label{sub:res_response}
 In order to further distinguish between MC and GC, we have also calculated the thermodynamic response functions of the system  (second derivatives of the thermodynamic potentials) like  baryon number susceptibility, the isothermal compressibility and the bulk modulus.
 The divergences/discontinuities of these response functions bear important signatures of phase transition. In Fig.~\ref{fig:3}(a, b and c ),  we have plotted the  three response functions ($\chi_N$, $K_T$,$K_B$) for neutron star and quark star in the same plot. For neutron star, $\chi_N$ (Fig.~\ref{fig:4}(a)) decreases very fast at low energy and then saturates and remain more or less constant as energy density increases. For quark star, $\chi_N$ increases with energy and then saturates depending on the stiffness of the EOS. In the next Fig.~\ref{fig:4}(b), we find that compressibility decreases with energy for both neutron star as well as quark star, the rate of decrease being very fast at lower energies. Bulk Modulus (Fig.~\ref{fig:4}(c)) which is just inverse of the isothermal compressibility shows a monotonic increase for both quark and neutron star. 
 
 In the next plot we show the results for hybrid stars using both Maxwell as well as the Gibbs construction. 
 We observe that $\chi_N$ diverges in the energy density gap region in  MC case, since $\chi_N \propto 1/C_s^2$ as $C_s^2$ equals zero there. In contrast, for the GC case, $\chi_N$ decreases initially and then rises again presenting a sharp peak where the mixed phase terminates and the pure quark phase start. This contrasting behavior provides an important signature of the nature of phase transition and charge neutrality, clearly differentiating MC from GC. It is also noteworthy that the  increase in the repulsive vector interaction reduces the baryon number susceptibility as can be seen from  Fig.~\ref{fig:4}(d). 
In Fig.~\ref{fig:4}(e), the isothermal compressibility $K_T$  has been plotted for both MC and GC and it can distinguish between both in the same way as the susceptibility. Since $K_T$ is related to $\chi_N$ by a factor of $1/\rho_B^2$, it exhibits a similar trend although the peak here at the start of mixed phase is more pronounced than that at the end of the mixed phase. 
In Fig.~\ref{fig:4}(f), bulk modulus has been plotted and since this is proportional to $C_s^2$, this distinguishes between MC and GC in the similar way as the speed of sound does. The peaks for GC are much more pronounced at the end of mixed phase where pure quark phase begins. In MC, $K_B$ becomes zero in the energy density gap.  The peaks of  $K_B$ during the onset of phase transition increases as $G_V$ increases in case of both GC and MC though the height of the peaks is much more in case of MC than GC. At the end of the phase transition too increase in $G_V$ results in increase in the height of the peak but the peaks in GC and MC here are comparable in magnitude.

Thus, beyond the commonly used sound velocity $C_s^2$, the quantities $\chi_N$, $K_T$ and $K_B$  emerge as valuable probes of the microscopic structure of the mixed phase. Their behavior not only characterizes the stability of the system but also provides an additional diagnostic tool to distinguish between a single conserved chemical potential (MC) and two conserved chemical potentials (GC). 
In the next Fig.~\ref{fig:5}, we have plotted
the three response functions and the speed of sound as a function of the mixed phase parameter $\chi$  (volume fraction in mixed phase \cite{Glendenning:1992vb}) for the Gibbs construction and the baryon number density $\rho_B$. The variables have been scaled appropriately in order to fit them in the same plot.   Fig.~\ref{fig:5} clearly displays the  relation between them in the mixed phase regime of the quark hadron phase transition in the core of compact stars. Bulk Modulus and the speed of sound has similar kind of variation in the mixed phase though the peaks are at different densities. Since $K_B$ is inversely proportional to $K_T$ , their maximum and minimum occur at the same value of $\chi$ or $\rho_B$ as expected.


\section{Neutron-star phenomenology}

\begin{figure}[htp] 
	\centering 
    \includegraphics[width=0.23\textwidth]{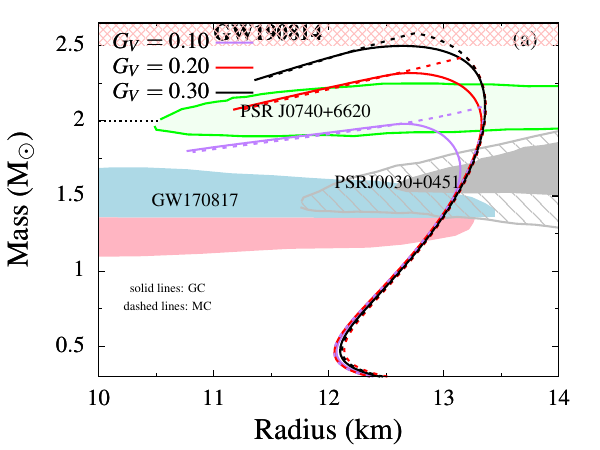}
    \includegraphics[width=0.23\textwidth]{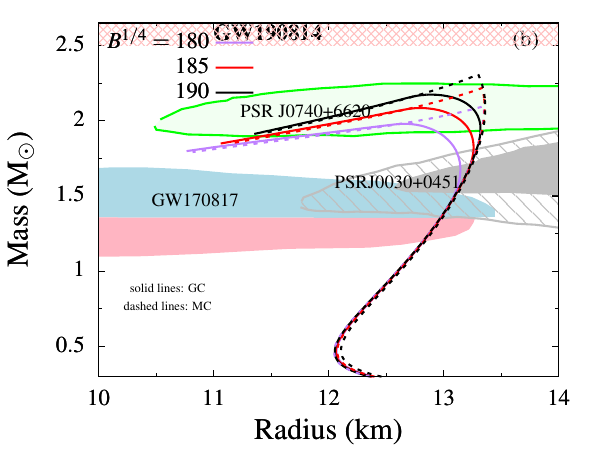}
        \includegraphics[width=0.23\textwidth]{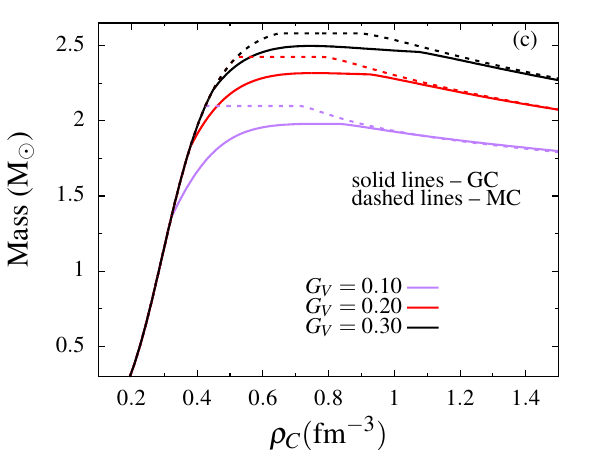}
    \includegraphics[width=0.23\textwidth]{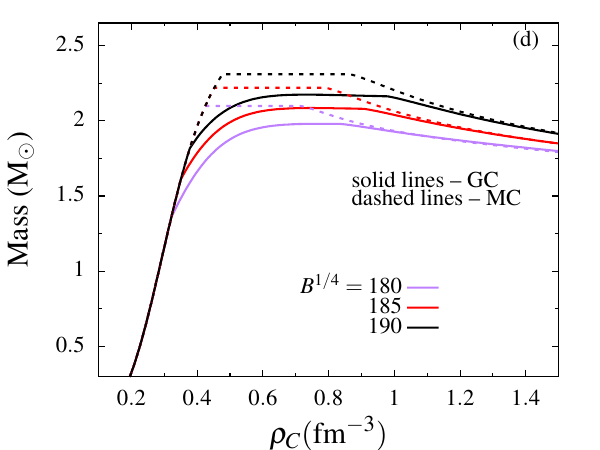}
\caption{Structural properties of hybrid stars obtained using the Gibbs (solid lines) and Maxwell (dashed lines) constructions. Panels (a) and (b) show the mass--radius relations for varying vector coupling strengths $G_V$ and bag constants $B$, respectively. Panels (c) and (d) display the stellar mass as a function of the central baryon density for the corresponding parameter sets. The shaded regions in the upper panels represent observational constraints from GW170817 \cite{LIGOScientific:2018cki}, PSR J0030+0451\cite{Choudhury:2024xbk}, and PSR J0740+6620\cite{Salmi:2024bss,Qi:2025mpn}, GW190814\cite{LIGOScientific:2020zkf}. }
\label{fig:mr_rhoc}
\end{figure}

To assess the implications of the present equations of state for neutron-star structure, we solve the Tolman--Oppenheimer--Volkoff (TOV) equations \cite{Tolman:1939jz,Oppenheimer:1939ne} and construct the corresponding stellar sequences. The resulting mass--radius relations and the dependence of stellar mass on the central baryon density are shown in Fig.~\ref{fig:mr_rhoc}, while the corresponding stellar properties are summarized in Table~\ref{tab:1}.
Figures~\ref{fig:mr_rhoc}(a) and (b) display the mass--radius relations for different values of the vector coupling $G_V$ and bag constant $B$, respectively. All considered equations of state satisfy the current observational constraints from PSR J0740+6620 and GW170817. The corresponding maximum masses lie in the range $M_{\rm max}\simeq 2.0$--$2.6\,M_\odot$, demonstrating that the hybrid equations of state remain sufficiently stiff at high densities.
The lower panels of Fig.~\ref{fig:mr_rhoc} show the stellar mass as a function of the central baryon density. These results provide a direct connection between the phase-transition densities obtained from the equation of state and the densities realized in neutron-star interiors. By comparing the transition densities with the central densities of stable stellar configurations, one can determine whether the mixed phase or pure quark matter is reached inside neutron stars. For the Gibbs construction considered here, the maximum-mass configurations contain a mixed hadron--quark phase but do not reach the pure quark phase. In contrast, the Maxwell construction allows the most massive stable stars to populate the post-transition branch associated with quark matter.

Table~\ref{tab:1} lists the maximum mass $M_{\rm max}$, corresponding radius $R_{\rm max}$, central density $\rho_c^{\rm max}$, central energy density $\epsilon_c^{\rm max}$, central pressure $P_c^{\rm max}$, and the canonical radii $R_{1.4}$ and $R_{2.0}$. The onset density of the phase transition, $\rho_{\rm onset}$, and the density associated with pure quark matter, $\rho_{\rm pureQ}$, are also reported. For brevity, Table~\ref{tab:1} reports representative stellar properties for the sequences obtained by varying $G_V$ at fixed $B$. The corresponding sequences obtained by varying $B$ exhibit qualitatively similar behavior in both the mass--radius relation and the phase structure, and are therefore shown only in Fig.~\ref{fig:mr_rhoc}(b). These quantities establish a direct link between the thermodynamic diagnostics discussed in this work and the composition of neutron-star cores.

\section{Summary and Conclusions}
\label{sec:Conclusions}
In this work, we have studied the decomposition of the speed of sound 
using three different schemes for pure hadron, pure quark and hybrid EOS 
using both MC and GC. One of the schemes emphasized in this work involves 
the average speed of sound $W$ and its logarithmic derivative $W^\prime$. 
The parameters of the other two decomposition methods 
($\alpha$, $\Delta$, $\beta$, $\Delta^{\prime}$) have been shown to be 
related to $W$ and $W^\prime$.  
The decomposition separates the averaged stiffness encoded in $W$ 
from its density evolution governed by $W'$. This distinction allows 
us to identify structural features of the equation of state that are 
not directly apparent from $c_s^2$ alone.
We find that in {\color{black}the pure hadronic phase, $W^{\prime}$ remains positive, ensuring that the averaged speed of sound never exceeds the local speed of sound. For self-bound quark matter, however, the generalized definition of the averaged speed of sound can lead to negative values of $W^{\prime}$ at low energy densities, followed by a sign change at higher densities. 
}In contrast, within the mixed phase, 
$W^{\prime}$ is always negative for MC and can be either positive or 
negative for GC. The polytropic index $\gamma$ is zero in case of MC and non-zero in GC 
for the mixed phase.
In order to gain deeper insights into the phase transition, we have also 
investigated other key thermodynamic quantities, such as the baryon number 
susceptibility, isothermal compressibility and the bulk modulus. 
The divergences of these response functions signal phase transitions. 
In the MC case, $\chi_N$ diverges in the energy density gap, while in the 
GC case it shows a sharp peak at the end of the mixed phase. Both $K_T$ 
and $K_B$ similarly distinguish MC from GC, with $K_B$ vanishing in MC 
but showing pronounced peaks in GC. Since $K_B \propto C_s^2$, it follows 
trends similar to the speed of sound, although their peaks occur at 
different densities. 

Finally, by solving the TOV equations, we established the connection between the thermodynamic diagnostics and neutron-star structure, showing that the mixed phase is realized in stable stellar configurations and that the appearance of pure quark matter depends sensitively on the adopted phase-transition construction.

\appendix

\section{Decomposition of the speed of sound in self-bound matter}
\label{app:selfbound}
For self-bound systems such as quark matter, the energy density at vanishing pressure is finite,
\begin{equation}
\varepsilon(P=0)=\varepsilon_0 \neq 0.
\end{equation}
This differs from hadronic matter, where the zero-pressure point is typically associated with vanishing energy density.

To define an averaged speed of sound, the natural reference point is the onset of nonzero pressure at $\varepsilon_0$. Accordingly, the averaged speed of sound can be written as
\begin{equation}
\langle c_s^2 \rangle
=
\frac{1}{\varepsilon-\varepsilon_0}
\int_{\varepsilon_0}^{\varepsilon}
c_s^2\, d\bar{\varepsilon}.
\end{equation}
Using the thermodynamic relation
\begin{equation}
c_s^2=\frac{dP}{d\varepsilon},
\end{equation}
together with $P(\varepsilon_0)=0$, one obtains
\begin{equation}
\langle c_s^2 \rangle
=
\frac{P}{\varepsilon-\varepsilon_0}.
\end{equation}

Motivated by this result, we define
\begin{equation}
W \equiv \frac{P}{\varepsilon-\varepsilon_0},
\end{equation}
which reduces to the standard expression $W=P/\varepsilon$ in the limit $\varepsilon_0\rightarrow0$. The pressure may then be expressed as
\begin{equation}
P=W(\varepsilon-\varepsilon_0).
\end{equation}
Differentiating with respect to $\varepsilon$ yields
\begin{equation}
c_s^2
=
W+(\varepsilon-\varepsilon_0)\frac{dW}{d\varepsilon}.
\end{equation}
Defining
\begin{equation}
W^\prime
\equiv
(\varepsilon-\varepsilon_0)\frac{dW}{d\varepsilon},
\end{equation}
the decomposition takes the familiar form
\begin{equation}
c_s^2 = W + W^\prime.
\end{equation}

Therefore, the structure of the decomposition remains unchanged for self-bound matter. The only modification is that the averaged quantity $W$ must be defined relative to the finite zero-pressure energy density $\varepsilon_0$. In the limit $\varepsilon_0\rightarrow0$, the standard hadronic-matter expressions are recovered.
\color{black}
\bibliography{hs}

\end{document}